\documentclass[aps,pra,twocolumn,showpacs,showkeys,amsmath,superscriptaddress,floatfix]{revtex4}

\usepackage{amsmath}
\usepackage{makeidx}
\usepackage{amsfonts}
\usepackage[utf8]{inputenc}
\usepackage[usenames,dvipsnames]{pstricks}
\usepackage{subfigure}
\usepackage{epsfig}
\usepackage{pst-grad} 
\usepackage{pst-plot} 
\usepackage[colorlinks,hyperindex]{hyperref}
\hypersetup
{
	colorlinks,%
	citecolor=black,%
	linkcolor=black,%
	urlcolor=black,%
}

\renewcommand{\vec}[1]{\mathbf{#1}}

\begin{document}
	
	\title{Higgs mode in the quench dynamics\\of a confined ultracold Fermi gas in the BCS regime}
	
	\author{S.~Hannibal}
	\affiliation{Institut f\"ur Festk\"orpertheorie, Westf\"alische
		Wilhelms-Universit\"at M\"unster, 48149 M\"unster,
		Germany}
	
	\author{P.~Kettmann}
	\affiliation{Institut f\"ur Festk\"orpertheorie, Westf\"alische
		Wilhelms-Universit\"at M\"unster, 48149 M\"unster,
		Germany}
	
	\author{M.~D.~Croitoru}
	\affiliation{Departement Fysica, Universiteit Antwerpen, 2020 Antwerpen, Belgium
	}
	
	\author{A.~Vagov}
	\affiliation{Theoretische Physik III, Universit\"at Bayreuth, 95440
		Bayreuth,
		Germany}
	
	\author{V.~M.~Axt}
	\affiliation{Theoretische Physik III, Universit\"at Bayreuth, 95440
		Bayreuth,
		Germany}
	
	\author{T.~Kuhn}
	\affiliation{Institut f\"ur Festk\"orpertheorie, Westf\"alische
		Wilhelms-Universit\"at M\"unster, 48149 M\"unster,
		Germany}
	
	\date{\today}
	
	\begin{abstract}
		The Higgs amplitude mode of the order parameter of an ultracold confined Fermi gas in
		the BCS regime after a quench of the coupling constant is analyzed
		theoretically. Characteristic features are a damped oscillation which at a
		certain transition time changes into a rather irregular dynamics. We
		compare the numerical solution of the full set of nonlinear equations of motion for
		the normal and anomalous Bogoliubov quasiparticle excitations with a linearized
		approximation. In doing so the transition time as well as the difference between resonant systems, i.e., systems where the Fermi energy is close to a subband minimum,
		and off-resonant systems can be well understood and traced back to the system
		and geometry parameters.
	\end{abstract}
	
	
	
	\maketitle
	
	\section{Introduction}
	
	Ultracold Fermi gases have been subject of many experimental and
	theoretical studies during recent years (see e.g. \cite{Bloch2008Many,Giorgini2008Theory,Yefsah2013Heavy,Tey2013Collective,Sidorenkov2013Second}). They provide a unique system to
	study key concepts of condensed matter theory. This is because in these
	systems many parameters such as the particle density, the Fermi energy, the
	confinement potential, or the interaction strength between the Fermions,
	which in a solid state system are typically fixed quantities, can be
	externally controlled in a wide range \cite{Chin2010Feshbach}. In particular, magnetic-field Feshbach
	resonances provide the means for controlling the interaction strength between
	fermions by varying an external magnetic field. The tunability of the s-wave
	scattering length, which is the dominant interaction channel, makes ultracold
	Fermi gases ideal for exploring different regimes of interacting many-body
	systems in a single system. This includes the limiting regimes of weakly
	attracting fermions, which condense into Cooper pairs forming a
	Bardeen-Cooper-Schrieffer (BCS) phase below a certain temperature $T_C$, and
	repulsive dimers formed by two fermions, which can undergo a Bose-Einstein
	condensation (BEC). These two limiting regimes are separated by the strongly
	interacting BCS-BEC crossover regime where the scattering length diverges and the system exhibits unitary properties \cite{Zwerger2011BCS}.
	
	In addition to the variable interaction strength, ultracold atomic gases
	offer a unique opportunity to explore the influence of a confinement on the
	pairing correlations, because dimensionality and confinement can be precisely
	controlled by tuning external parameters
	\cite{Bloch2008Many,Martikainen2005Quasi,Dyke2011Crossover,Shanenko2012Atypical}.
	Varying the confinement, which is often well approximated by a harmonic
	confinement potential, allows one to access new degrees of freedom.
	Restricting the Fermi gases to quasi-low dimensionality may, e.g., offer the
	possibility for an experimental evidence of unconventional phases, like the
	Fulde-Ferrell-Larkin-Ovchinnikov (FFLO) state \cite{Fulde1964Superconductivity,Larkin1972Influence,Yonezawa2008Anomalous,Tomadin2008Nonequilibrium,Devreese2011Controlling,Croitoru2012In}. Moreover it may
	help to study and get experimental insight into shape resonances,
	theoretically predicted for quasi-low dimensional conventional
	superconductors \cite{Shanenko2006Size}. In Ref. \cite{Dyke2011Crossover} the first quantitative measurements of the
	transition from 2D to quasi-2D and 3D in a weakly interacting Fermi gas has been reported. At
	low atom numbers, the shell structure associated with the
	filling of individual transverse oscillator states has been observed. On the theoretical side
	the ground state properties of a $^{6}$Li gas confined in a cigar-shaped
	laser trap have been investigated predicting size-dependent resonances of the
	superfluid gap \cite{Shanenko2012Atypical}, similar to the case of
	superconducting nanowires \cite{Shanenko2006Size}, yielding an atypical BCS-BEC crossover.
	
	An important effort is now devoted to the exploration of the
	out-of-equilibrium behavior of trapped ultracold atomic Fermi gases and,
	in particular, to the determination of their dynamical properties. The dynamics has been studied in the normal as well as in the condensed phase, observing second sound \cite{Sidorenkov2013Second} and soliton trains \cite{Yefsah2013Heavy}, and showing a low-frequency oscillation of the cloud after a change of the system confinement or optical excitation \cite{Guajardo2013Higher,Riedl2008Collective,Bartenstein2004Collective,Pantel2012Trap,Lepers2010Numerical}. Furthermore, state-of-the-art technology allows one to change the coupling constant on such short time scales that it is possible to explore the regime where the many-body system is governed by a unitary evolution with nonequilibrium
	initial conditions. In ultracold atomic Fermi gases the dynamics may be initiated by readjusting the pairing interaction through switching an external magnetic field in the region of a Feshbach resonance (i.e., a quantum quench) or by a rapid change of the confinement potential of the trap \cite{Riedl2008Collective}. Due to the small energies in the trapping potential the dynamics in the Fermi gases take place on a millisecond timescale. Therefore, in contrast to metallic superconductors, where sub-picosecond excitations are required to achieve non-adiabatic dynamics \cite{Papenkort2007Coherent}, in atomic gases the non-adiabatic regime can be reached already by excitations in the sub-millisecond range.
	
	Spontaneous symmetry breaking gives rise to collective modes of the order parameter which are classified into the Higgs amplitude mode, and the Goldstone mode, the latter corresponding to a phase oscillation of the gap \cite{Nambu1960Axial,Varma2002Higgs,Volovik2014Higgs}. The Higgs mode has been subject of intensive theoretical and experimental \cite{Sooryakumar1980Raman,Littlewood1981Gauge,Matsunaga2014Light} research efforts in the past. On the theoretical side the non-adiabatic temporal response of the order parameter to (quasi-)instantaneous perturbations has been studied. Different regimes of an oscillatory temporal behavior of the pairing potential were theoretically predicted in homogeneous fermionic condensates \cite{Barankov2004Collective,Barankov2006Synchronization,Yuzbashyan2006Relaxation,Dzero2007Spectroscopic}.
	It was shown that the amplitude of the order parameter oscillates without damping when the coupling constant is increased above a certain critical value
	\cite{Barankov2006Synchronization}. On the other hand, the gap vanishes when the
	coupling constant is decreased below another critical value. In between these
	two limiting regimes the amplitude exhibits damped dephased
	oscillations and the system goes to a stationary steady state with a finite
	gap \cite{Barankov2006Synchronization}. In extended systems the approach to a
	stationary state occurs in an oscillatory way with an inverse square root
	decay in time of the amplitude of the oscillations \cite{Yuzbashyan2006Relaxation}. A similar evolution was predicted for conventional bulk superconductors
	\cite{Papenkort2007Coherent,Papenkort2008Coherent} where the non-adiabatic regime is reached by excitation with short, intense terahertz pulses. An experimental realization was reported in \cite{Matsunaga2013Higgs}. In contrast, in finite length superconducting nanowires a breakdown of the damped oscillation and a subsequently rather irregular dynamics has been predicted
	\cite{Zachmann2013Ultrafast}.
	
	In this paper we present a theoretical analysis of the short-time BCS
	dynamics of a $^{6}$Li gas confined in a cigar-shaped laser trap. The
	excitation is modeled by a sudden change of the interaction strength which
	can be achieved through a Feshbach resonance by an abrupt change of the
	external magnetic field \cite{Chin2010Feshbach}. Applying the well-known BCS
	theory in mean field approximation to ultracold Fermi gases we show that the
	change of the coupling strength induces a collective oscillation of the
	Bogoliubov quasiparticles close to the Fermi level. This results in a damped amplitude
	oscillation of the BCS gap, which corresponds to the Higgs mode \footnote{In the case of a nonlinear system given here the Higgs and the Goldstone mode are in general coupled. However, due to the considered weak excitations and thus weak dynamical coupling the system behaves approximately linear and the Higgs and the Goldstone mode decouple.}. Like in the case of confined BCS superconductors this oscillation breaks down after a certain time revealing rather chaotic dynamics afterwards. We explain these dynamics in terms of coupled harmonic oscillators which can be derived by linearizing the quasiparticle dynamics obtained from the Heisenberg equation of motion.
	
	In doing so we first derive the quasiparticle equations of motion from the
	inhomogeneous Bogoliubov-de Gennes Hamiltonian (Sec.~\ref{sec:formalism})
	which, because of using the standard contact-type interaction, requires a
	proper regularization of the gap equation. Starting from the ground state
	calculated according to Ref.~\cite{Shanenko2012Atypical} we then calculate the dynamics
	of the superfluid gap after an instantaneous change of the coupling constant.
	The numerical results as well as their explanation follow in
	Sec.~\ref{sec:results}, where we first discuss a rather small system and then
	proceed to a larger, experimentally more easily accessible system. Finally,
	in Sec.~\ref{sec:conclusions} we summarize our results and give some
	concluding remarks.

	\section{Theoretical approach}\label{sec:formalism}
	
	Our approach aims at describing the dynamics of the superfluid order
	parameter $\Delta(\vec{r},t)$ of an ultracold $^{6}$Li Fermi gas, confined in
	a cigar-shaped, axial symmetric harmonic trapping potential
	\begin{equation}
	V_\text{conf}(x,y,z) = \frac{1}{2} m \omega_\perp^2 (x^2+ y^2) +
	\frac{1}{2} m \omega_\parallel^2 z^2. \label{eq:Vconf}
	\end{equation}
	Here, $m$ is the mass of the $^{6}$Li atoms and $\omega_\perp$
	($\omega_\parallel$) is the confinement frequency in the $x$-$y$-plane
	($z$-direction), respectively. Choosing $\omega_\perp \gg \omega_\parallel$
	yields an elongated cigar-shaped trap where the oscillator length $l_\alpha =
	\sqrt{\hbar / (m \omega_\alpha)}$ provides a measure of the system length.
	The eigenvalues  
	\begin{equation}\label{eq:oneparticle}
	\varepsilon_m = \hbar \omega_\perp(m_x + m_y + 1) + \hbar\omega_\parallel(m_z+\frac{1}{2}) - E_F
	\end{equation} are measured with respect to the Fermi
	energy $E_F$. The index $m$ refers to the combination of quantum numbers
	$m_x$, $m_y$, and $m_z$. For this geometry the one-particle states form
	one-dimensional subbands, labeled by $(m_x,m_y)$ [cf. Fig.~\ref{fig:states}],
	while the states within each subband are labeled by $m_z$. Each subband has a
	constant one-particle density of states and thus the overall density of
	states exhibits finite jumps whenever a new subband appears.
	
	We consider the gas to be composed of two spin states, $\uparrow$ and
	$\downarrow$, and start from the inhomogeneous BCS Hamiltonian at $T=0\,
	\text{K}$. Within the Anderson approximation we then derive equations of
	motion for the corresponding Bogoliubov quasiparticle excitations.

	\subsection{Hamiltonian}
	
	The usual inhomogeneous BCS Hamiltonian for an effective BCS-type contact
	interaction reads \cite{DeGennes1989Superconductivity}
	\begin{align}\label{eq:BCS-H}
	H_{BCS} &= \int \, \left[ \Psi_{\uparrow}^{\dagger}(\vec{r}) H \Psi_{\uparrow}^{}(\vec{r})
	+ \Psi_{\downarrow}^{\dagger}(\vec{r}) H \Psi_{\downarrow}(\vec{r}) \right] \, d^3 r\notag \\
	&- g \int   \, \Psi_{\uparrow}^{\dagger}(\vec{r}) \Psi_{\downarrow}^{\dagger}(\vec{r})
	\Psi_{\downarrow}^{}(\vec{r})  \Psi_{\uparrow}^{}(\vec{r}) \, d^3 r,
	\end{align}
	where $\Psi_{\uparrow}(\vec{r})$ and $\Psi_{\downarrow}(\vec{r})$ are the
	field operators for up and down spin, respectively, $H = p^2 / 2 m +
	V_{\text{conf}}-E_F$ is the one-particle Hamiltonian and $g$ is the coupling
	constant of the contact interaction $V(\vec{r}) = - g\,\delta(\vec{r})$. In
	the limit of low temperatures the main contribution to the interaction
	between two fermionic atoms in different internal spin states is given by
	scattering processes at low momentum. The description of those can be
	replaced by the widely known pseudopotential only depending on the scattering
	length $a$ \cite{Bloch2008Many}, which yields $ g = - \frac{4 \pi \hbar^2
		a}{m}$.
	
	A BCS-like mean field expansion in terms of anomalous expectation values and
	a particle-hole transformation, leaving spin-up operators unchanged,
	$\Psi_{\uparrow}^\dagger = \Phi_{\uparrow}^\dagger$, while interchanging
	spin-down ones, $\Psi_\downarrow^\dagger = \Phi_\downarrow$, leads to the
	Bogoliubov-de Gennes (BdG) Hamiltonian
	\cite{Datta1999Can}
	\begin{align}\label{eq:BDG}
	H_{BdG} &= \int \, \Phi_{\uparrow}^{\dagger}(\vec{r}) H \Phi_{\uparrow}^{}(\vec{r})
	\,  d^3 r - \int \,   \Phi_{\downarrow}^{\dagger}(\vec{r}) H \Phi_{\downarrow}(\vec{r})
	\,d^3 r \notag \\
	&+  \int   \, \left( \Delta(\vec{r}) \Phi_{\uparrow}^{\dagger}(\vec{r})
	\Phi_{\downarrow}^{}(\vec{r}) + \Delta^*(\vec{r}) \Phi_{\downarrow}^{\dagger}(\vec{r})
	\Phi _{\uparrow}^{}(\vec{r}) \right) \, d^3 r,
	\end{align}
	where \begin{equation} \Delta(\vec{r}) = - g \left<
	\Psi_{\downarrow}(\vec{r}) \Psi_{\uparrow}(\vec{r}) \right>= - g \left<
	\Phi_{\downarrow}^{\dagger}(\vec{r}) \Phi_{\uparrow}^{}(\vec{r}) \right>
	\end{equation}
	is
	the BCS order parameter. From Eq. \eqref{eq:BDG} it becomes apparent that the corresponding eigenvalue equation can be
	written as the one-particle Bogoliubov-de Gennes equation
	\begin{equation}\label{eq:Bogoliubov-Gl}
	\begin{pmatrix} H & \Delta(\vec{r}) \\ \Delta^*(\vec{r}) & -H^* \end{pmatrix}
	\left(\begin{array}{c} u_M(\vec{r}) \\ v_M(\vec{r}) \end{array} \right)
	= E_M \left( \begin{array}{c} u_M(\vec{r}) \\ v_M(\vec{r}) \end{array} \right).
	\end{equation}
	$H_{BdG}$ can thus be diagonalized by Bogoliubov's transformation, using
	the eigenfunctions $u_M(\vec{r})$ and $v_M(\vec{r})$. This introduces
	non-interacting quasiparticles with energy $E_M$ obeying fermionic
	commutation relations with the corresponding creation operator
	\begin{equation}\label{eq:B-Transformation}
	\gamma_M^{\dagger} = \int \, \left[  u_M(\vec{r}) \Phi_{\uparrow}^{\dagger}(\vec{r})
	+ v_M(\vec{r}) \Phi_{\downarrow}^{\dagger}(\vec{r}) \right] \, d^3 r.
	\end{equation}
	
	The spectrum of the BdG equation is symmetric with respect to the Fermi
	energy and thus the eigenstates of the BdG equation occur in pairs. Labeling
	$M \rightarrow m,\alpha$ for states $E_M > 0$ and $M \rightarrow m,\beta$ for
	$E_M< 0$, respectively, one finds the relations $u_{m,\beta} = -
	v_{m,\alpha}^*$ and $v_{m,\beta} = u_{m,\alpha}^*$ for the eigenstates.
	Therefore, all quantities can be expressed solely using the $\alpha$ state wave functions
	\cite{Datta1999Can}. In the following we omit the $\alpha$,  $\beta$ index of
	the eigenfunctions while they are still necessary for the quasiparticle operators. Hereafter --in the case of the eigenfunctions-- the index $m$ refers to the $\alpha$ states.
	For our further calculations it is convenient to transform to the excitation
	picture ($\alpha \rightarrow a$, $\beta \rightarrow b$) with
	$\gamma_{m\alpha} = \gamma_{ma}$ and $\gamma_{m\beta} =
	\gamma_{mb}^{\dagger}$, where all quasiparticle excitations vanish in the
	ground state. We can rewrite the order parameter in the basis given by the
	eigenfunctions, where it reads:
	\begin{align}\label{eq:Delta_Bogolon}
	\Delta(\vec{r},t) = - g  \sum_{m,n}  & v_m^*(\vec{r}) u_n(\vec{r}) \Big<
	\gamma_{ma}^{\dagger}\gamma_{na}^{\phantom{\dagger}}\Big> \notag \\
	+ & u_m(\vec{r}) u_n(\vec{r}) \left< \gamma_{mb}^{\phantom{\dagger}}
	\gamma_{na}^{\phantom{\dagger}} \right> \notag \\
	- & v_m^*(\vec{r}) v_n^*(\vec{r}) \left< \gamma_{ma}^{\dagger}\gamma_{nb}^{\dagger}
	\right> \notag \\
	+ & u_m(\vec{r}) v_n^*(\vec{r}) \left( \left< \gamma_{nb}^{\dagger}\gamma_{mb}^{}
	\right> -\delta_{mn}\right).
	\end{align}
	This yields the well-known result for the ground state order parameter
	\begin{equation}\label{eq:Delta0}
	\Delta_{GS}(\vec{r}) =  g \sum_{m} u_m(\vec{r}) v_m^*(\vec{r}),
	\end{equation}
	which has to be solved self-consistently with Eq.~\eqref{eq:Bogoliubov-Gl}
	\cite{DeGennes1989Superconductivity}. Focusing on the underlying physics, we
	exploit the Anderson approximation (A.A.) \cite{Anderson1959Theory}, choosing the
	BdG wave functions $u_m(\vec{r})$ and $v_m(\vec{r})$ proportional to the
	one-particle wave functions of the confinement potential
	$\varphi_m(\vec{r})$, i.e.,
	\begin{equation}\label{eq:Anderson}
	u_m(\vec{r}) = u_m \varphi_m(\vec{r}) \mbox{\quad} \text{and} \mbox{\quad}
	v_m(\vec{r}) = v_m \varphi_m(\vec{r}).
	\end{equation}
	Here the amplitudes of the BdG wave functions $u_m$, $v_m$ are obtained from the BdG Eq.~\eqref{eq:Bogoliubov-Gl} and read
	\begin{equation}
		u_m = \sqrt{\frac{1}{2} \left(1+ \frac{\varepsilon_m}{E_m}\right)} \qquad  v_m = \sqrt{\frac{1}{2} \left(1 - \frac{\varepsilon_m}{E_m}\right)}
	\end{equation}
	with the quasiparticle energies given by
	\begin{equation}
	E_m = \sqrt{\varepsilon_m^2 + \Delta_m^2}
	\end{equation}
	and the one-particle energies $\varepsilon_m$ given by Eq. \eqref{eq:oneparticle}. Applying the Anderson approximation to Eq.~\eqref{eq:Bogoliubov-Gl} additionally yields $\Delta_{mn} = \left<m\right|
	\Delta(\vec{r}) \left| n \right> = \Delta_{m} \delta_{mn}$, where $\left| m
	\right>$ are the one-particle eigenfunctions. The Anderson approximation has
	been tested in several nanostructured geometries and no qualitative
	deviations have been found \cite{Chen2009Superconducting}. It is applied to
	all our calculations.
	
	From Eqs.~\eqref{eq:Bogoliubov-Gl} and \eqref{eq:Delta0} we obtain the
	well-known BCS-like self-consistency equation, also referred to as gap
	equation. The ground state order parameter in the state $\left| m \right>$ is
	given by
	\begin{equation}\label{eq:Selbstkonsistenz}
	\Delta_m =  - \sum_{m'} V_{mm'} \frac{\Delta_{m'}}{2 E_{m'}},
	\end{equation}
	Here $V_{mm'}$ is the interaction matrix element
	\begin{equation}\label{eq:Wechselwirkungsmatrix1}
	V_{mm'} = - g \int  \, \left| \varphi_m(\vec{r})\right|^2 \left|
	\varphi_{m'}(\vec{r})\right|^2 d^3 r,
	\end{equation}
	which exhibits maxima for states at the subband minimum   (i.e., states with
	low $m_z$).
	
	The contact interaction used here leads to a well-know ultraviolet divergence
	in the summation over all states, i.e., in Eq.~\eqref{eq:Selbstkonsistenz},
	which can be regularized by applying a scattering length regularization
	\cite{Bloch2008Many}. This has been established for the homogeneous gap
	equation and subsequently extended to the inhomogeneous gap equation
	\eqref{eq:Selbstkonsistenz}, where Ref. \cite{Bruun1999BCS} gives a careful derivation for confined systems. However, Refs. \cite{Bruun2002Cooper,Shanenko2012Atypical} state that a much simpler regularization is sufficient since the results are not sensitive to the details of the method used. The corresponding regularized gap equation reads
	
	\begin{equation}\label{eq:Selbstkonsistenz-reg}
	\Delta_m =  -  \frac{1}{2} \, \sum_{m'} V_{mm'} \Delta_{m'} \left( \frac{1}{E_{m'}}
	- \frac{1}{\varepsilon_{m'} + E_\text{F}} \right),
	\end{equation}
	which can be rewritten as a multiplication by a factor
	\begin{equation}\label{eq:reg-Faktor-dyn}
	\chi_{m'} = \left( 1 - \frac{E_{m'}}{\varepsilon_{m'} + E_\text{F}} \right).
	\end{equation}
	Thus, each quasiparticle state $m'$ is weighted by a factor $\chi_{m'}$. 
	In order to extend this in a consistent manner to the nonequilibrium case, in
	which $\Delta$ deviates from its ground state value $\Delta_{GS}$ and is
	determined by the full Eq.~\eqref{eq:Delta_Bogolon}, the same procedure has to be applied to the nonequilibrium version of Eq.~\eqref{eq:Selbstkonsistenz-reg} as will be discussed below [see Eq. \eqref{eq:Delta-DeltaT}] \footnote{This regularization introduces an additional dependency on the quasiparticle energy $E_m$ and thus the ground state gap $\Delta_m$ in the dynamical equations. Since --in the dynamical case-- the time-dependent gap $\Delta_m(t)$ differs from the ground state value this may seem unusual. However, we have checked that the method is not sensitive on the detailed parameters used, e.g., on using one constant value for all $\Delta_m$ in order to omit the newly introduced dependence. Thus the proposed regularization scheme is a suitable extension since the physical behavior of the system remains unchanged.}.
	
	\begin{figure*}[ht!]
		\subfigure[]{
			\label{fig:states}
			\includegraphics[width = 0.17\textwidth]{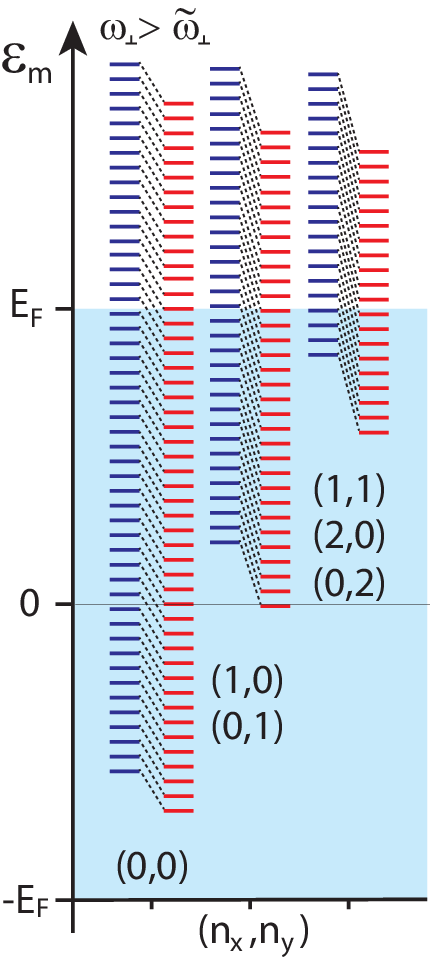}
		}
		\subfigure[]{
			\label{fig:dispersion}
			\includegraphics[width = 0.36\textwidth]{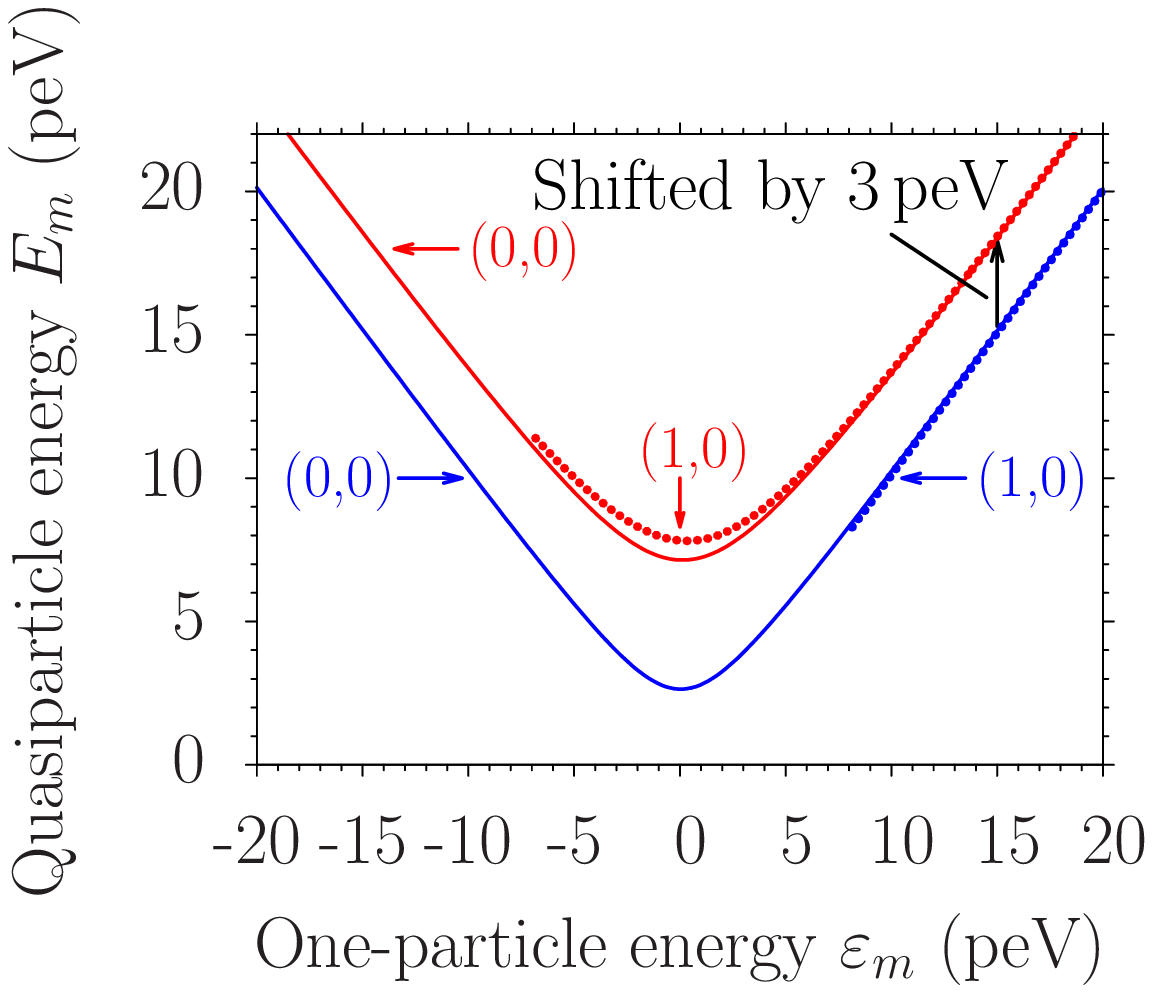}
		}
		\subfigure[]{
			\label{fig:densityofstates}
			\includegraphics[width = 0.36\textwidth]{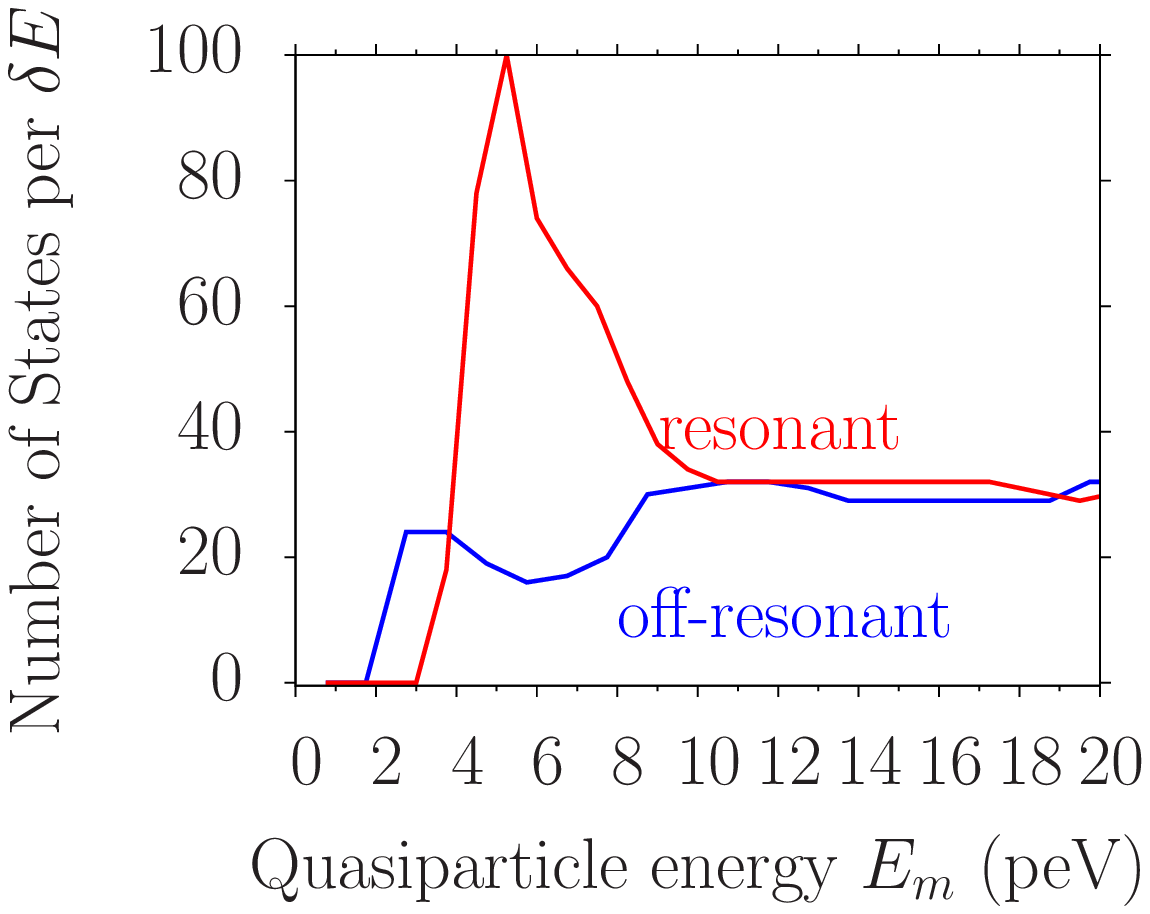}
		}
		\caption{(Color online) One-particle and quasiparticle properties of a resonant system
			(red) and an off-resonant system (blue). (a) Schematic one-particle states for
			two differently sized systems (i.e., different $\omega_\perp$) (b) quasiparticle energy vs. one-particle energy, solid lines: subband $(0,0)$, dashed lines: subbands
			$(1,0)$, $(0,1)$ (higher subbands are not visible), subbands of resonant system are shifted up by $3\,$peV for better visibility,  and (c) quasiparticle
			density of states.}
	\end{figure*}
	
	Figure~\ref{fig:dispersion} shows the dependence of the quasiparticle energies
	$E_m$ on the one-particle energies $\varepsilon_m$. For all subbands crossing
	the Fermi energy finite minima of the quasiparticle energy evolve at
	$\varepsilon = 0$. Subbands with the minimum close to the Fermi energy (i.e.,
	$\varepsilon_{m_x,m_y,m_z=0} \approx 0$) exhibit a larger $\Delta_m$ due to
	the larger interaction matrix elements $V_{mm'}$. This leads to a rather
	shallow parabolic-like minimum in these subbands and, thus, to a maximum in
	the corresponding density of states (see Fig.~\ref{fig:densityofstates}).
	Subbands showing this feature will be called resonant in the following and
	systems in which such a subband exists are referred to as resonant systems.
	The behavior can be compared to the well-known parabolic-like dispersion
	relation in the homogeneous BCS theory with $\Delta = \text{const}$. A
	similar behavior has also been obtained for nanostructured superconducting
	systems \cite{Shanenko2010Giant}. Overall the density of states combined with
	the interaction matrix $V_{mm'}$ leads to the quantum size oscillations of
	the order parameter upon changing the lateral size of the system (i.e.,
	$\omega_\perp$) found in Ref.~\cite{Shanenko2012Atypical}.

	\subsection{Dynamics}
	
	In this paper we consider an excitation of the Fermi gas by a quantum quench, i.e., a sudden change of the coupling constant $\tilde{g} \rightarrow g$. Since in the region of a Feshbach resonance the coupling constant strongly depends on an external
	magnetic field $\vec{B}$, this can be achieved experimentally by rapidly
	switching $\vec{B}$ from an initial value $\vec{B}_i$ to a final value
	$\vec{B}_f$. We assume that this switching process occurs on a time scale
	much faster than the characteristic time scale of the order parameter
	dynamics, such that the excitation can be taken to be instantaneous. This
	assumption is realistic since fast linear magnetic ramps with rates of
	$240\,$G/ms are experimentally available \cite{Chin2006Evidence} while the
	timescale of the gap dynamics is of the order of $\tau_\Delta \approx \hbar /
	\Delta_{GS} \sim 1\,$ms \cite{Barankov2004Collective}. The excitations
	considered in this paper require a shift in the magnetic field of a few
	gauss, which indeed can be assumed to be instantaneous on the typical ms time
	scale in ultracold Fermi gases. In this case during the switching of the
	magnetic field the state of the system remains unchanged.
	
	As usual, the dynamics of a quantum mechanical system can be described in
	different basis systems, which from a mathematical point of view are all
	equivalent. In our case, to calculate the dynamics of the order parameter
	after a quench from the \textit{initial} system
	($\tilde{u}_m(\vec{r})$, $\tilde{v}_m(\vec{r})$, $\tilde{g}$) to the
	\textit{final} system ($u_m(\vec{r})$, $v_m(\vec{r})$, $g$) we choose a
	time-independent basis rather than remaining in the diagonal basis. For
	convenience we take the basis corresponding to the eigenstates of the system
	after the switching, i.e., to the coupling constant $g$. All our calculations
	are thus carried out in the basis $u_m(\vec{r})$, $v_m(\vec{r})$.
	
	The initial state, which is characterized by the ground state order parameter
	$\tilde{\Delta}_{GS}(\vec{r})$, corresponding to the coupling constant
	$\tilde{g}$ and the basis functions $\tilde{u}_m(\vec{r})$,
	$\tilde{v}_m(\vec{r})$, has to be expressed in terms of the basis
	$u_m(\vec{r})$, $v_m(\vec{r})$, which in particular gives rise to
	non-vanishing quasiparticle excitations in this basis. Since the confinement potential is unchanged, also the corresponding one-particle wave
	functions $\varphi_m(\vec{r})$ remain unchanged. According to
	Eq.~\eqref{eq:Anderson}, in the present case only the BdG amplitudes $u_m$
	and $v_m$ change while the spatial shapes of $u_m(\vec{r})$ and
	$v_m(\vec{r})$ remain unchanged. Therefore, all orthogonality relations are
	preserved and only diagonal quasiparticles are populated. For the initial values of the normal and anomalous expectation values, respectively, one finds
	\begin{align}\label{eq:Anregung_Kopplung}
	\Big<\gamma_{ma}^{\dagger}\gamma_{ma}^{}\Big>\big|_{t=0}
	&= \left(v_m \tilde{u}_m - u_m  \tilde{v}_m\right)^2 \\
	\left<\gamma_{ma}^{\dagger}\gamma_{mb}^{\dagger}\right>\big|_{t=0}
	&= \left(v_m \tilde{u}_m - u_m \tilde{v}_m \right) \left(v_m \tilde{v}_m + u_m \tilde{u}_m\right).
	\end{align}
	In addition
	\begin{equation}
	\Big<\gamma_{ma}^{\dagger}\gamma_{ma}^{\phantom{\dagger}}\Big> =
	\left<\gamma_{mb}^{\dagger}\gamma_{mb}^{\phantom{\dagger}}\right> \, , \,
	\left<\gamma_{mb}^{\phantom{\dagger}}\gamma_{ma}^{\phantom{\dagger}}\right> =
	\left<\gamma_{ma}^{\dagger}\gamma_{mb}^{\dagger}\right>^*
	\end{equation}
	holds for all times $t \geq 0$.

	Since the instantaneous order parameter $\Delta(t)$ deviates from the ground
	state value of the final system $\Delta_{GS}$, the Hamiltonian in the basis
	$u_m(\vec{r})$, $v_m(\vec{r})$ becomes non-diagonal depending on the
	difference $(\Delta(t) - \Delta_{GS})$. It thus becomes explicitly time
	dependent according to
	\begin{align}\label{eq:BdG-Hamilton}
	\lefteqn{H_{BdG} = \sum_m E_{ma} \gamma_{ma}^{\dagger} \gamma_{ma}
		- E_{mb}\left(1-\gamma_{mb}^{\dagger}\gamma_{mb}\right)} \notag \\
	&+ \sum_{m,n} \left[\left(\Delta-\Delta_{GS}\right)_{u_m^* v_n}
	+ \left(\Delta^*-\Delta_{GS}^*\right)_{v_m^* u_n} \right]
	\gamma_{ma}^{\dagger}\gamma_{na}^{} \notag \\
	&+ \sum_{m,n} \left[\left(\Delta-\Delta_{GS}\right)_{u_m^* u_n^*}
	- \left(\Delta^*-\Delta_{GS}^*\right)_{v_m^* v_n^*} \right]
	\gamma_{ma}^{\dagger}\gamma_{nb}^{\dagger} \notag \\
	&- \sum_{m,n} \Big[\left(\Delta-\Delta_{GS}\right)_{v_m v_n} -
	\left(\Delta^*-\Delta_{GS}^*\right)_{u_m u_n} \Big]
	\gamma_{mb}^{}\gamma_{na}^{} \notag \\
	&- \sum_{m,n}  \left[\left(\Delta-\Delta_{GS}\right)_{v_m u_n^*}
	+ \left(\Delta^*-\Delta_{GS}^*\right)_{u_m v_n^*} \right]
	\left(1-\gamma_{mb}^{\dagger}\gamma_{nb}^{}\right).
	\end{align}
	with
	\begin{align}
	\left(\Delta - \Delta_{GS} \right)_{u_m^* v_n^{}} & = \int
	\: u_m^*(\vec{r}) \left[\Delta(\vec{r},t) - \Delta_{GS}(\vec{r}) \right]
	v_n(\vec{r}) d^3 r \notag \\
	& \underbrace{=}_{\text{A.A.}} \left(\Delta - \Delta_{GS} \right)_{m}
	u_m^{} v_m^{} \delta_{mn}^{}.
	\end{align}
	Here, the Anderson approximation has been applied to the dynamical equations
	as proposed in Ref.~\cite{Zachmann2013Ultrafast}, yielding $\left(\Delta -
	\Delta_{GS} \right)_{mn} \approx \left(\Delta - \Delta_{GS} \right)_{m}
	\delta_{mn}^{}$. The time evolution of the system is thus described by the
	time-dependent quasiparticle expectation values. The corresponding equations
	of motion can be obtained via Heisenberg's equation of motion.
	
	For the considered instantaneous change of the coupling constant only diagonal expectation values are excited. The required equations of motion read
	\begin{align}
	i \hbar \frac{d}{dt} \Big<\gamma_{ma}^{\dagger}\gamma_{ma}^{\phantom{\dagger}}\Big>
	&= a_m \left<\gamma_{ma}^{\dagger}\gamma_{mb}^{\dagger}\right>^* - a_m^*
	\left<\gamma_{ma}^{\dagger}\gamma_{mb}^{\dagger}\right> \label{eq:Bwg-Gl-normal}\\
	i\hbar \frac{d}{dt}  \left<\gamma_{ma}^{\dagger}\gamma_{mb}^{\dagger}\right>
	& = -2\, E_m^{\text{(ren)}}  \left<\gamma_{ma}^{\dagger}\gamma_{mb}^{\dagger}\right> \notag \\
	&\quad \qquad + a_m \left( 1 - 2 \Big<\gamma_{ma}^{\dagger}\gamma_{ma}^{\phantom{\dagger}}\Big>
	\right) \label{eq:Bwg-Gl-anomal},
	\end{align}
	where
	\begin{equation}
	E_m^{\text{(ren)}} = E_m + 2 \, u_m v_m \rm{Re}
	\left[(\Delta-\Delta_{GS})_m\right], \label{eq:E_ren}
	\end{equation}
	\begin{equation}
	a_m = v_m^2(\Delta-\Delta_{GS})_m - u_m^2(\Delta-\Delta_{GS})_m^* , \label{eq:a_m}
	\end{equation}
	and
	\begin{align}\label{eq:Delta-DeltaT}
	(\Delta - \Delta_{GS})_{m} &=
	-  \sum_{k} \Bigg[ 2 v_k u_k \left< \gamma_{ka}^{\dagger}\gamma_{ka}^{\phantom{\dagger}}
	\right>  \notag \\
	& + u_k^2 \left< \gamma_{kb}^{\phantom{\dagger}}\gamma_{ka}^{\phantom{\dagger}} \right>
	- v_k^2 \left< \gamma_{ka}^{\dagger}\gamma_{kb}^{\dagger} \right> \Bigg] V_{mk} \chi_k.
	\end{align}
	In Eq.~\eqref{eq:Delta-DeltaT} again the regularization factor $\chi_k$ has
	been introduced. Equations \eqref{eq:Bwg-Gl-normal}-\eqref{eq:Delta-DeltaT}
	represent a finite set of coupled ordinary differential equations that we
	solve numerically.
	
	It is interesting to note that the evolution of the anomalous expectation
	values [Eq.~\eqref{eq:Bwg-Gl-anomal}] corresponds to a set of harmonic
	oscillators with energies approximately given by $2 E_m$
	[first term in Eq.~\eqref{eq:Bwg-Gl-anomal}] while
	Eq.~\eqref{eq:Bwg-Gl-normal} as well as the second terms of
	Eqs.~\eqref{eq:Bwg-Gl-anomal} and \eqref{eq:E_ren} contain nonlinear couplings to all other
	oscillators via the factor $(\Delta-\Delta_{GS})_m$, which vanishes when the order parameter
	agrees with its ground state value. We will come back to this separation into
	linear and nonlinear terms in Sec.~\ref{sec:linearized}.

	\section{Results}\label{sec:results}
	
	In the following the temporal evolution of the amplitude of the spatially averaged gap
	\begin{equation}
	\bar{\Delta} = \frac{1}{V}\int d^3r \Delta(\vec{r})
	\end{equation}
	will be shown and analyzed for different system parameters. Here, the normalization
	volume $V$ is set to $V=l_x l_y l_z$ with $l_\alpha$ being the oscillator length
	in $\alpha$ direction \footnote{We want to remark that this definition is only for
		illustrative purposes. The normalization only scales the
		gap dynamics but does not enter the dynamical equations.}.
	
	In order to
	concentrate on the physics we will start our analysis by investigating a very
	small system: All the main features occurring in the dynamics of larger,
	experimentally accessible confinements arise in small systems, too, but with
	a strongly reduced degree of numerical complexity. Thus, the dynamics of the
	superfluid gap will at first be explained on the basis of small systems. The
	results for a larger system will be shown afterwards.
	
	\subsection{Full model, small system}
	
\begin{figure}[b]
		\includegraphics[width=1\columnwidth]{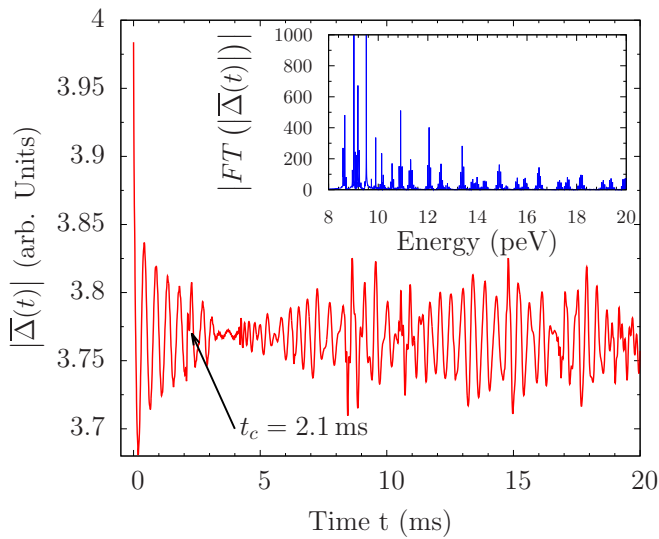}
		\caption{(Color online) Dynamics of the spatially averaged gap after a sudden change of the
			scattering length from $-140\,$nm to $-135\,$nm; inset: Fourier transform of the
			gap dynamics. The confinement frequencies are $f_\perp = 11.2\,$kHz
			and $f_\parallel = 240\,$Hz.}
		\label{fig:dynamik+ft}
	\end{figure}

	An exemplary result for the gap dynamics after a quantum quench, obtained by changing the scattering length from
	$\tilde{a}=-140\,$nm to $a=-135\,$nm for a system with the confinement
	frequencies $f_\perp = \omega_\perp/2\pi = 11.2\,$kHz and $f_\parallel =
	\omega_\parallel/2\pi = 240\,$Hz, is shown in Fig.~\ref{fig:dynamik+ft}. The
	Fermi energy has been set to $E_F = 100\,\hbar\omega_\parallel$ yielding $1/(k_F a) \approx -1$ according to Ref.~\cite{Shanenko2012Atypical}. As can be seen the amplitude of the gap shows an initial
	drop corresponding to the decreased coupling and thus decreased ground state
	gap. Afterwards a smoothly damped oscillation around the new ground state
	value of the gap occurs, which after a certain transition time $t_c$ turns
	into an irregular, rather chaotic oscillation. Here $t_c$ is defined as the
	time of the first deviation \footnote{In the numerical analysis the
		transition time $t_c$ is assumed to be reached if the distance between two
		adjacent maxima shows a deviation of more then 50$\%$ compared to the
		averaged distance of all preceding maxima.} from a smooth oscillation.
	
	The inset of Fig.~\ref{fig:dynamik+ft} suggests that this irregular
	oscillation after $t_c$ is the result of a superposition of several
	frequencies. Here a segment of the Fourier spectrum of the gap dynamics is
	shown. The spectrum is
	\begin{figure}[t]
		\centering
		\includegraphics[width=1\columnwidth]{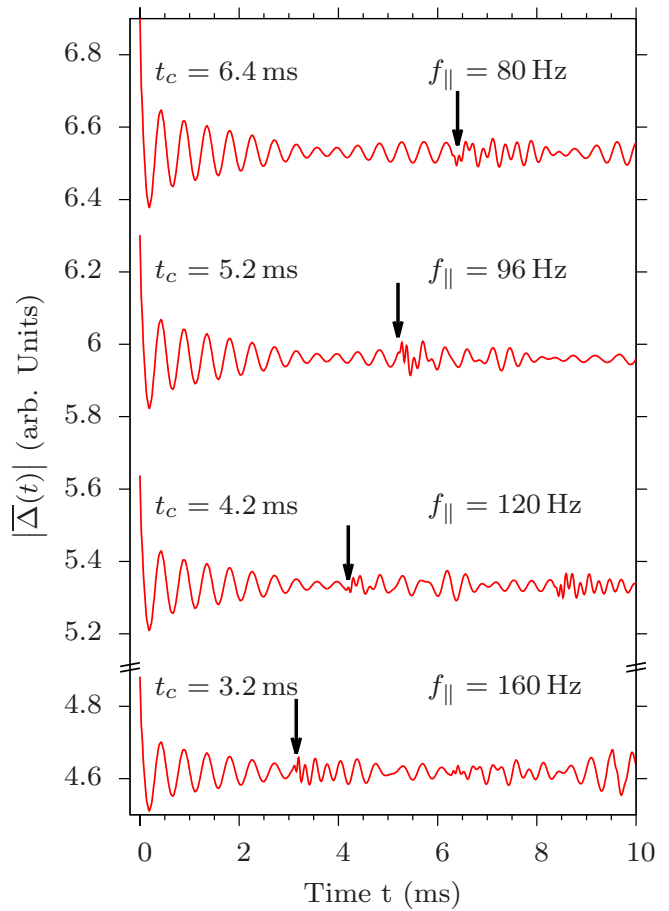}
		\caption{(Color online) Dynamics of the spatially averaged gap showing the change of the transition time (i.e., the time of the first deviation
			from a smooth oscillation) for a decreasing parallel confinement frequency
			(from bottom to top). The perpendicular confinement frequency is $f_\perp = 11.2\,$kHz.
			The arrows mark the transition time $t_c$.}
		\label{fig:uebergangszeit_entwicklung}
	\end{figure}
	composed of a series of sharp peaks in the range of $8.9\,$peV to about
	$30\,$peV, which can each be assigned to a corresponding quasiparticle state,
	i.e., $\hbar \omega_m \approx 2 E_m$. The main peaks at the lower end of this
	series correspond to the frequency of the initial damped oscillation and to
	the dominant frequencies of the irregular dynamics afterwards. Their values
	are given by the quasiparticle energies closest to the Fermi level. These lie
	in the vicinity of a quasiparticle subband minimum. The corresponding
	frequencies are thus given by $\hbar \omega_m \approx
	2\Delta_{(m_x,m_y,m_\text{min})}$, where $m_\text{min}$ is the $z$ quantum
	number referring to the state with minimal quasiparticle energy, i.e., the
	state at the Fermi energy. The other peaks belong to higher quasiparticle
	states and decrease continuously with increasing energy.
	
	While the qualitative picture of the gap dynamics is the same for all
	investigated systems, the quantitative values of the features mentioned above
	crucially depend on the system parameters. On the one hand the ground state
	gap and thus the mean value of the oscillation and its frequency
	contributions strongly depend on the perpendicular confinement $f_\perp$ (due to the
	size-dependent superfluid resonances \cite{Shanenko2012Atypical}) and on the
	scattering length $a$. The transition time, on the other hand, increases with
	decreasing parallel confinement $f_\parallel $ --i.e., with increasing system
	length-- as can be seen in Fig.~\ref{fig:uebergangszeit_entwicklung}. Here
	the gap dynamics is shown for the same perpendicular confinement and
	excitation as in Fig.~\ref{fig:dynamik+ft} but for increasing system length,
	i.e., decreasing $f_\parallel$ (from bottom to top). Figure
	\ref{fig:uebergangszeit_entwicklung} shows that the transition time $t_c$
	moves to larger times as the length of the system increases. In addition a
	revival of the oscillation can be seen for the two largest systems with
	$f_\parallel = 96\,$Hz and $f_\parallel = 80\,$Hz, which for smaller systems
	would occur after the breakdown.
	
\begin{figure}[t]
		\includegraphics[width=1\columnwidth]{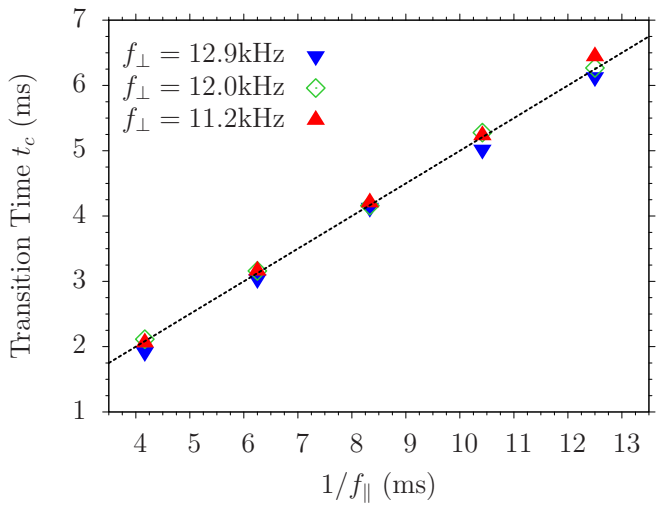}
		\caption{(Color online) Transition times of three different perpendicular confinements in
			dependence of the inverse parallel confinement frequency $\frac{1}{f_\parallel} \sim
			l_z^2$ compared to the prediction from the linearized theory (dashed curve).}
		\label{fig:uebergangszeit}
	\end{figure}

	A quantitative analysis of the transition time for different perpendicular
	confinements $f_\perp$ over a wide range of parallel confinements is shown in
	Fig.~\ref{fig:uebergangszeit}. Here the transition time is plotted against
	the inverse parallel confinement frequency $\frac{1}{f_\parallel} \sim
	l_z^2$. One can see that $t_c$ is independent of the size of the gas in the
	$x$-$y$-plane, since the values for every $f_\perp$ lie on the same curve. It
	is only influenced by the confinement in $z$-direction, where a linear
	dependence on $\frac{1}{f_\parallel}$ can be observed. This is in full
	agreement with the behavior found for superconducting quantum wires
	\cite{Zachmann2013Ultrafast}. Even a beating-like pattern was found for thin
	quantum wires which corresponds to the revivals seen in
	Fig.~\ref{fig:uebergangszeit_entwicklung}.
	
\begin{figure}[t]
		\centering
		\includegraphics[width=1\columnwidth]{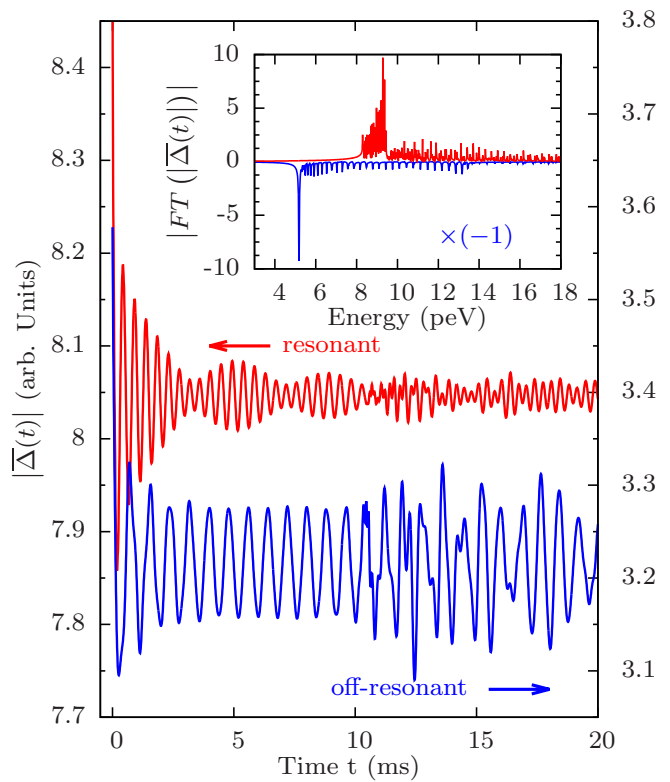}
		\caption{(Color online) Dynamics of the averaged gap for a resonant ($f_\perp = 11.2\,$kHz;
			red curve) and an off-resonant system ($f_\perp = 12.9\,$kHz; blue curve); in
			both cases $f_\parallel = 48\,$Hz. Inset: Fourier spectra.}
		\label{fig:resonant_vergleich}
	\end{figure}

	To investigate the smooth regime of the gap dynamics
	Fig.~\ref{fig:resonant_vergleich} shows calculations for a rather large
	system length and two different perpendicular confinements. For this case of
	large lengths in Ref.~\cite{Zachmann2013Ultrafast} it was found that thick
	quantum wires exhibit a damping of the gap oscillation given by a power law
	$\sim t^{-\alpha}$ with $\alpha=3/4$ for resonant systems and $\alpha = 1/2$
	for off-resonant ones. However, thin quantum wires were found to differ from
	this power law showing a more irregular oscillation but still a rather fast
	decay of the gap oscillation when resonant subbands are present.
	
	Figure \ref{fig:resonant_vergleich} shows that this situation applies to
	ultracold Fermi gases as well. Here a calculation of the gap dynamics is
	shown for a resonant system (upper, red curve), which is again characterized by the same
	perpendicular confinement as in Figs.~\ref{fig:dynamik+ft} and
	\ref{fig:uebergangszeit_entwicklung}, as well as for a system far away from
	resonance (lower, blue curve; $f_\perp = 12.9\,$kHz). The excitation is the same as
	before and the parallel confinement frequency is chosen as $f_\parallel =
	48\,$Hz, which corresponds to a rather long cloud. It can be seen that both
	systems show an initial decay of the gap oscillation until a minimal
	amplitude is reached. In the resonant case this initial decay is rather
	strong and fast. Here, the amplitude of the oscillation falls close to zero.
	Afterwards it exhibits revivals until the smooth oscillation breaks down. In
	contrast, the off-resonant system shows an only moderate, comparatively slow
	decay of the oscillation, which after a short time exhibits a nearly constant
	amplitude. Thus, on the one hand the decay of the oscillation is much
	stronger in the resonant than in the off-resonant case. On the other hand
	revivals and a beating like pattern occur for the resonant case before the
	breakdown while systems far away from resonance exhibit a nearly constant
	oscillation amplitude. The inset of Fig.~\ref{fig:resonant_vergleich} shows
	that these different temporal evolutions correspond to different Fourier
	spectra. Here the Fourier transforms of the resonant (positive $y$-axis) and
	the off-resonant (negative $y$-axis) system are shown. The resonant spectrum
	is composed of several strong components at the lower end and weaker peaks
	towards higher energies. In contrast, the off-resonant spectrum --it is
	shifted due to the smaller gap-- contains only one dominant frequency part at
	low energies while the rest of the spectrum is strongly suppressed. Thus,
	resonant systems on the one hand correspond to spectra with several strong
	modes and weaker high-frequency parts. Off-resonant systems on the other hand
	are strongly dominated by one single mode with only minor contributions from
	other energies.

	The features described above --the damped oscillation, the breakdown, and the
	irregular dynamics-- will be explained in the following section. As mentioned
	before, they also occur in BCS-superconductors in a very similar way, where
	cigar-shaped Fermi gases correspond to thin, short quantum wires. The
	following explanations therefore apply to both ultracold Fermi gases and
	confined BCS-superconductors thus showing the close relation between these
	systems.
	
	\subsection{Linearized dynamics, small system}\label{sec:linearized}
	
	To analyze the mechanisms underlying the gap dynamics and its features we
	introduce a linearized set of equations of motion. This can be derived by
	neglecting all terms of second and higher order in the quasiparticle
	excitations (this is valid due to the weak excitation investigated in this
	paper, which leads to $\left|\langle\gamma^\dag_{m a}\gamma_{m
		b}\rangle\right| \ll \left|\langle\gamma^\dag_{m a}\gamma^\dag_{m
		b}\rangle\right| \ll 1$). In doing so, Eq.~\eqref{eq:Bwg-Gl-normal} can be
	neglected since by inserting Eqs.~\eqref{eq:a_m} and \eqref{eq:Delta-DeltaT}
	into this equation only terms of at least second order in the normal
	excitations or products of anomalous and normal excitations contribute, which
	are to be neglected in the linearized case.
	
	\begin{figure}[b]
		\centering
		\includegraphics[width=1\columnwidth]{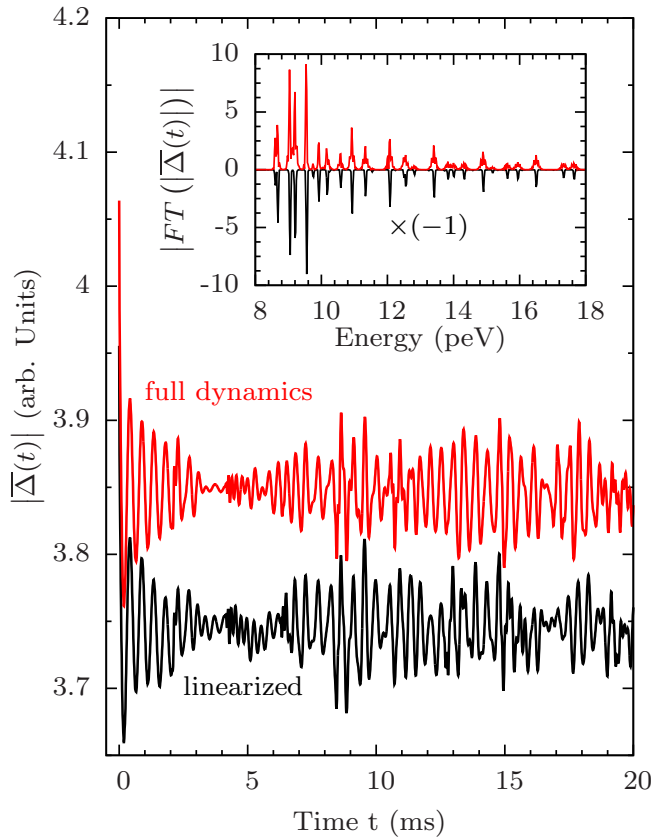}
		\caption{(Color online) Gap dynamics derived from the linearized equation of motion (black)
			compared to the full dynamics (red); inset: Fourier transforms. The
			confinement frequencies are $f_\perp = 11.2\,$kHz and $f_\parallel = 240\,$Hz.
		}
		\label{fig:linear_vergleich}
	\end{figure}
	
	Equation \eqref{eq:Bwg-Gl-anomal} reduces to a closed set
	of equations for the anomalous excitations which, by performing the
	derivative in time of one equation and inserting the other, can be separated
	into its real and imaginary parts leading to
	\begin{eqnarray}
	\dfrac{d^2}{dt^2}\langle\gamma^\dag_{m a}\gamma^\dag_{m b}\rangle +
	\omega_m^2 \langle \gamma^\dag_{m a}\gamma^\dag_{m b}\rangle &=& \nonumber \\
	\sum_{k\neq m} \big[ A_{km}  \text{Re}(\langle \gamma^\dag_{k a}\gamma^\dag_{k b}\rangle)
	+ i A_{mk} \hspace{-0.2cm}&\text{Im}&\hspace{-0.2cm}(\langle\gamma^\dag_{k a}\gamma^\dag_{k b}\rangle)
	\big]\chi_k, \hspace{0.01cm} \label{eq:Bwg-Gl-lin}
	\end{eqnarray}
	where again all terms nonlinear in the quasiparticle expectation values have
	been neglected. This equation describes a set of linearly coupled harmonic
	oscillators with the uncoupled frequencies
	\begin{equation}
	\omega_m = \sqrt{ \left(\dfrac{2E_m}{\hbar}\right)^2 - A_{mm}\chi_m } \label{eq:Omega-lin}
	\end{equation}
	and
	\begin{equation}
	A_{km} = \dfrac{2}{\hbar^2} V_{km} \left(E_k + \dfrac{\varepsilon_m
		\varepsilon_k}{E_k} \right) - \frac{1}{\hbar^2} \sum_{l} \frac{\varepsilon_k}{E_k}
	\frac{\varepsilon_l}{E_l} V_{ml} V_{lk}\chi_l. \label{eq:Akm-lin}
	\end{equation}
	The coupling strength of the oscillators $A_{km}$ with $k\neq m$ therein is
	weak due to the --in this case-- mostly small matrix elements $V_{km}$. The
	shift of the eigenfrequencies of the coupled system
	[Eq.~\eqref{eq:Bwg-Gl-lin}] with respect to the uncoupled frequencies
	[Eq.~\eqref{eq:Omega-lin}] should therefore be small, as should be the shift
	of the uncoupled frequencies compared to the bare ones $2 E_m/\hbar$.

	Figure \ref{fig:linear_vergleich} shows the dynamics of the BCS gap obtained
	from Eq.~\eqref{eq:Bwg-Gl-lin} compared to the full dynamics \footnote{In
		case of possible divergences of the linearized solution one can artificially
		lower the coupling strength in the dynamical calculations in order to prevent
		them. This means that while the ground state is still calculated with the
		coupling $g$, in the dynamical calculations (i.e., in Eqs.~\eqref{eq:Bwg-Gl-lin},\eqref{eq:Omega-lin} and \eqref{eq:Akm-lin}) the coupling has to be slightly lowered by a small amount, i.e., $g \rightarrow g^\prime \lesssim g$. It should be mentioned that the full (nonlinear) equations never lead to
		divergences.}. The parameters correspond to the system shown in
	Fig.~\ref{fig:dynamik+ft} and are exemplary for all investigated systems. The
	linearized equations clearly reproduce the full dynamics and all its features
	in very good agreement. The inset shows that the positions as well as the
	strengths of most of the frequencies in the Fourier spectra are well
	described by this approximation. The spectrum corresponding to the full
	equations of motion (upper, red curve) and the one corresponding to the linearized
	equation (lower, black curve) show only slight differences. Only one low lying weak
	frequency component close to zero, which is present in the linearized version
	(outside the range shown in the inset of Fig.~\ref{fig:linear_vergleich}) as well as weak side
	peaks that occur adjacent to every main peak in the full dynamics are not
	fully reproduced. The latter can be attributed to the nonlinear couplings.
	Their influence on the dynamics, however, is obviously negligible.
	
	The main features of the gap dynamics can thus be explained on the basis of
	Eq.~\eqref{eq:Bwg-Gl-lin}, which can be solved analytically. The analytic
	solution can be expressed in terms of a linear superposition of simple
	(co)sine-oscillators. The corresponding frequencies are determined by the
	uncoupled oscillator frequencies $\omega_m$ and the coupling $A_{km}$. They are
	thus completely fixed by the the system parameters and the BCS gap, but they
	are independent of the initial conditions. In the considered case of weak
	coupling the coupled spectrum is only slightly shifted compared to the uncoupled
	frequencies. The eigenfrequencies of the coupled dynamics are thus
	approximately given by twice the quasiparticle energies, as observed above.
	
	The amplitudes of the different eigenmodes of the coupled system are determined by the initial values of the dynamics and thus depend on the details of the excitation. In general, one observes that each of the quasiparticle oscillators $\langle\gamma^\dag_{m a}\gamma^\dag_{mb}\rangle$ carries strong contributions from oscillators in the vicinity of its uncoupled frequency $\omega_m$ and in areas of a high density of states. These areas are located close to the minimum of a quasiparticle subband (see section
	\ref{sec:formalism}), i.e., near $\hbar \omega = 2 \Delta_{(m_x,m_y,m_\text{min})}$. In addition one observes that the contributions from each oscillator to the lowest energies are mostly in phase while these to higher energies are more and more out of phase. A sum of all
	quasiparticle oscillators, which according to Eq.~\eqref{eq:Delta_Bogolon}
	yields the BCS gap, thus leads to dynamics dominated by low-lying frequencies
	at a quasiparticle band minimum. These can be understood as cumulative peaks
	created by collective oscillations of all quasiparticles in the system.
	
	All spectra obtained by the full equations indeed show such a spectrum with
	rather dense, strong frequency contributions near $\hbar \omega = 2
	\Delta_{(m_x,m_y,m_\text{min})}$ and relatively widespread suppressed
	frequency contributions at higher energies.
	
	The explanations given above have shown that the gap dynamics can be
	understood as a linear superposition of quasiparticle oscillations which
	themselves are given by a sum of simple oscillators. To finally explain the
	main features of the gap dynamics in the time domain --the damping, the
	transition and the irregular oscillation-- one can therefore use an even
	simpler picture of a set of independent cosine-oscillators with the
	frequencies $\omega_m$ \footnote{The main features can actually be reproduced
		by plotting a weighted sum of such cosine functions.}.
	
	Due to the abrupt excitation all these oscillators start in phase at maximum
	deflection. Starting to oscillate they will soon dephase. A sum of all
	oscillators (i.e., the gap) thus performs a damped oscillation (cf.
	Ref.~\cite{Yuzbashyan2006Relaxation}). Here systems with several strong
	frequency contributions --i.e., resonant systems-- show a rather fast and
	persistent damping since a large part of the cumulative amplitude is able to
	dephase. Off-resonant systems in contrast exhibit only a slight decay of the
	oscillation since the main part of the cumulative oscillation is carried by
	one single mode.
	
	Proceeding in time the damped oscillation continues until all oscillators are
	completely dephased and the amplitude of the oscillation is minimal. Then the
	oscillators start to rephase, the amplitude grows and the oscillation
	reappears (see Fig.~\ref{fig:uebergangszeit_entwicklung}; for off-resonant
	systems this effect is strongly suppressed since one single frequency
	dominates the spectrum). As soon as the first adjacent pair of oscillators
	goes back in phase again this beating-like pattern is interrupted. At this
	moment a spike in the cumulative oscillation indicates the breakdown of the
	regular oscillation and thus the transition time $t_c$. Afterwards all other
	frequencies rephase successively and create a rapid sequence of spikes which
	leaves a picture of an irregular oscillation.

	The preceding argumentation suggests that the time of this breakdown should
	be inversely proportional to the maximum spacing of adjacent quasiparticle
	energies. Strictly speaking the transition times are found to be determined
	by adjacent quasiparticles from the same subband
	\cite{Zachmann2013Ultrafast}, i.e.,
	\begin{equation}
	\hspace*{-4mm}t_c \hspace*{-0.5mm} \approx \hspace*{-0.5mm} \frac{2\pi\hbar}{2 \delta E_{\text{max}}}
	\hspace*{-0.5mm} \approx \hspace*{-0.5mm} \frac{\pi\hbar}{(\varepsilon_{m_x,m_y,m_z+1}
		\hspace*{-0.5mm}-\varepsilon_{m_x,m_y,m_z})} = \frac{1}{2f_\parallel}.\hspace*{-2mm} \label{eq:TE}
	\end{equation}
	It thus increases $\sim 1/f_\parallel$ with decreasing parallel confinement
	frequency since the energy spacing of the atomic spectrum then decreases.
	This is in full agreement with the relation found before. In fact,
	Eq.~\eqref{eq:TE} gives exactly the linear curve shown in
	Fig.~\ref{fig:uebergangszeit}. This indicates that the breakdown of the
	smooth initial oscillation of the BCS gap indeed is due to adjacent
	frequencies rephasing in time.
	
	\subsection{Full model, large system}
	
\begin{figure}[t]
		\hspace*{-1cm}
		\includegraphics[width=1\columnwidth]{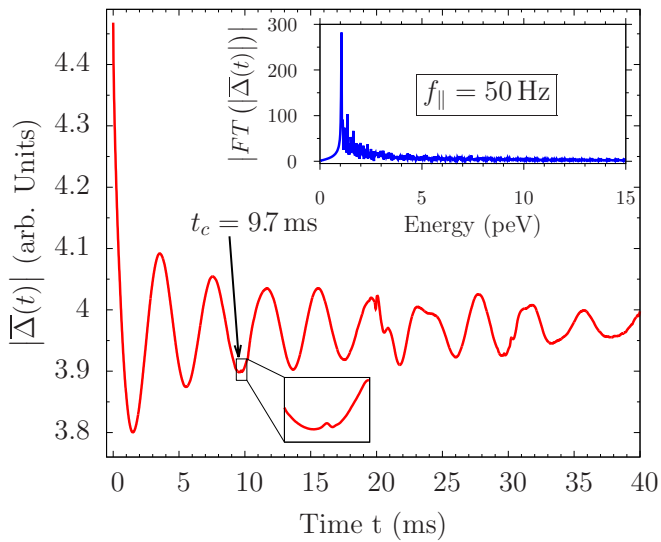}
		\caption{(Color online) Gap dynamics for a large system after a sudden change
			of the scattering length from $-140\,$nm to $-135\,$nm; inset: Fourier
			transform of the gap dynamics. The confinement frequencies are
			$f_\perp = 1.03\,$kHz and $f_\parallel = 50\,$Hz.}
		\label{fig:dynamik_gross}
	\end{figure}

	As a final example the gap dynamics of a rather large system with
	$f_\parallel =50\,$Hz and $f_\perp \approx 1.03\,$kHz again after a sudden
	change of the scattering length from $-140\,$nm to $-135\,$nm is shown in
	Fig.~\ref{fig:dynamik_gross}. This corresponds to a gas with $l_\parallel
	\approx 5.4\,\mathrm{\mu m}$ and $l_\perp \approx 1.2\,\mathrm{\mu m}$ and is
	thus on an experimentally accessible length-scale \cite{Bartenstein2004Collective}. The Fermi
	energy is chosen as $E_F = 250 \hbar\omega_\parallel$ which corresponds to
	$N=12258$ atoms in the trap.
	
	Figure \ref{fig:dynamik_gross} shows that the qualitative behavior of the gap
	dynamics is the same as for the smaller systems: A slowly decaying
	oscillation of the gap occurs. The transition time $t_c = 10\,\mathrm{ms}$
	calculated from Eq.~\eqref{eq:TE} is in good agreement with a small bump in
	the curve at $t\approx9.7\,$ms, the first deviation from a smooth
	oscillation. Afterwards more and more deviations occur and the gap dynamics
	becomes successively irregular.
	
	When looking at the damping of the gap oscillation one can see that although
	the system is resonant --one subband is close to the Fermi energy-- the
	strength of the damping is situated somewhere between the resonant and the
	off-resonant case of Fig. \ref{fig:resonant_vergleich}. This is due to the
	weak perpendicular confinement and therefore large number of states
	contributing to the condensate: Compared to the overall number of relevant
	states the resonant ones give only a small contribution to the gap. For
	larger systems the resonances are thus less pronounced
	\cite{Shanenko2012Atypical}.
	
	The Fourier transform in the inset of Fig.~\ref{fig:dynamik_gross} shows a
	familiar picture, too, with strong contributions at the lower end of the
	spectrum and successively decaying peaks towards higher energies. The
	difference with respect to the spectra shown before is on the one hand the
	high density of peaks for the large system. This is due to the weaker
	confinement and thus higher density of bare and quasiparticle states. On the
	other hand the gap in the Fourier spectrum and thus the main frequency of the
	oscillation is comparatively small. This is because of the larger ratio of
	the perpendicular to the parallel length $l_\perp/l_\parallel$. With the
	Fermi energy fixed at $E_F = 250 \hbar \omega_\parallel$ this leads to a
	comparatively low particle density of the trapped gas and thus a weaker
	condensate and a smaller gap.
	
	\section{Conclusion}\label{sec:conclusions}
	
	In conclusion, we have calculated the Higgs amplitude dynamics in the BCS phase of an ultracold $^6$Li gas confined in a cigar-shaped trap. The dynamics is induced by a quantum quench resulting from a sudden change of an external magnetic field. We have shown that the amplitude of the spatially averaged gap performs a damped oscillation breaking down after a certain time $t_c$, which is determined by the parallel confinement frequency $f_\parallel$, i.e., by the length of the cloud. Afterwards a rather irregular oscillation involving many different frequencies occurs.
	
	We have investigated the influence of the confinement on the gap dynamics and
	the impact of the size-dependent superfluid resonances on its qualitative
	behavior. It turned out that in the case of a resonant system, i.e., a system
	where the Fermi energy is close to a subband minimum, the dynamics of the
	order parameter exhibits a strong damping and, for sufficiently long systems,
	a revival before eventually the irregular regime is reached. In contrast, in
	an off-resonant system the damping is much less pronounced and the oscillation
	before the transition to the irregular regime is mainly determined by a
	single frequency.
	
	By analyzing the linearized version of the equations of motion for the
	quasiparticle excitations we were able to interpret the observed features of
	the dynamics. It turned out that for the excitations studied in this paper
	the linearized equations well reproduce the dynamical behavior of the gap,
	except for some slight details resulting from the nonlinearities in the full
	equations of motion. From the linearized model it becomes evident that the
	system approximately behaves like a set of weakly coupled harmonic
	oscillators. The frequencies as well as the couplings of these oscillators
	are completely determined by the system parameters after the excitation while
	the amplitudes of the different eigenmodes depend on the details of the
	excitation. The analysis revealed in particular that the transition time to
	the irregular dynamics is directly related to the energy separation of the
	one-particle energies while the differences between resonant and non-resonant
	systems is caused by the different densities of states and coupling
	efficiencies close to the subband minima.

	\begin{acknowledgments}
		M.D.C. acknowledges the support by the BELSPO Back to Belgium Grant.
	\end{acknowledgments}


\end{document}